\documentclass[12pt,preprint]{aastex}

\shorttitle{YOUNG SOLAR-TYPE STARS}
\shortauthors{Biazzo et al.}

\begin{document}

\title{PHOTOSPHERIC AND CHROMOSPHERIC ACTIVE REGIONS IN FOUR YOUNG SOLAR-TYPE 
STARS\altaffilmark{1}}

\author{K. BIAZZO and A. FRASCA}
\affil{INAF - Catania Astrophysical Observatory, via S. Sofia 78, 95123 Catania, Italy}
\email{kbiazzo@oact.inaf.it, afr@oact.inaf.it}

\author{G. W. HENRY}
\affil{Center of Excellence in Information Systems, Tennessee State University,
3500 John A. Merritt Blvd., Box 9501, Nashville, TN 37209}

\and

\author{S. CATALANO and E. MARILLI}
\affil{INAF - Catania Astrophysical Observatory, via S. Sofia 78, 
95123 Catania, Italy}

\altaffiltext{1}{Based on observations collected at the {\it Osservatorio Astrofisico di Catania} (Italy) and Fairborn Observatory (USA).}

\begin{abstract}
We present a photometric and spectroscopic study of four G--K dwarfs, namely 
HD~166, $\epsilon$~Eri, $\chi^1$~Ori and $\kappa^1$~Cet.  In three cases, we 
find a clear spatial association between photospheric and chromospheric 
active regions.  For $\chi^1$~Ori we do not find appreciable variations of 
photospheric temperature, and chromospheric H$\alpha$ emission.  We applied 
a spot/plage model to the observed rotational modulation of temperature and 
flux to derive spot/plage parameters and to reconstruct a rough 
``three-dimensional'' map of the outer atmosphere of $\kappa^1$~Cet,
HD~166 and $\epsilon$~Eri.
\end{abstract}


\section{INTRODUCTION}

The atmospheres of main-sequence (MS) stars with spectral types later than 
F5 exhibit the effects of magnetic activity.  At ages of a few hundred 
Myr, MS stars have magnetic activity levels higher than the Sun but 
considerably lower than what is typically observed in close binary RS CVn 
systems or in BY Dra stars.  Systematic photometric monitoring of MS FGK 
stars has been pursued for more than three decades to study starspots in 
these stars (see, e.g., \citealt{Radick83, Radick98}; \citealt{Lockwood97};
\citealt{Hen99}).  In particular, \cite{Radick83} reported that two out of 
eleven stars monitored in Str\"omgren $uvby$ passbands showed light 
variations anti-correlated with the contemporaneous \ion{Ca}{2} H\&K S-index. 
\cite{Lockwood97} found small-amplitude variations in $b$ and $y$ filters for 
about ten of 41 stars in their Lowell Observatory program, monitored for about 
eleven years, and reported photometric variability correlated with mean 
chromospheric activity.  In a more recent paper, \cite{Radick98} analyzed 
the same dataset and found short-term variability behavior due to rotational 
modulation in at least 15 of the solar-like stars in their sample. 


Recently, several studies have been conducted to analyze the photospheric 
and chromospheric active regions in young stars (\citealt{Stra93}; 
\citealt{Stou99}), but all the objects studied in these works are ultra-fast 
rotators (UFR) or pre-main sequence (PMS) stars.  In particular, \cite{Stra93} 
found a marginal correlation between the starspot distribution and 
chromospheric inhomogeneities in LQ Hya, a rapidly rotating 
($P_{\rm rot}\simeq1\fd6$) single K2 V star, probably just arriving on the 
Zero Age Main Sequence (ZAMS).  However, they could not discriminate between plages and local velocity 
fields as the cause of the observed variations in the Full-Width Half-Maximum 
(FWHM) of the H$\alpha$ line.  On the other hand, the two very active rapidly 
rotating Pleiades stars HII 686 and HII 3163 studied by \cite{Stou99} did 
not display the maximum of H$\alpha$ and \ion{Ca}{2} Infrared Triplet (IRT) 
emission at the same phase of the spot transit.  A rotational modulation of 
the H$\alpha$ emission with maximum nearly coincident with the minimum of 
the light curve has been instead detected by \cite{Frasca97} in the active 
rapidly rotating binary TZ CrB.  Recently, \cite{Frasca00} presented the 
first photospheric/chromospheric study of HD 206860, a solar-type star with 
an activity level intermediate between the Sun and very active PMS and UFR 
stars; they used \ion{Ca}{2} H\&K and H$\alpha$ lines as chromospheric 
indicators and photometric observations as a photospheric indicator.  They 
found a clear rotational modulation in all the chromospheric and photospheric 
indicators, proving the presence of an uneven distribution of long-lived 
active regions and a spatial association between photospheric spots and 
chromospheric plages, as displayed in the Sun and in the most active 
RS CVn systems.

In this paper, we study the photospheric and chromospheric surface 
inhomogeneities in four young MS stars with activity levels intermediate 
between the Sun and the very active UFR stars.  We use light curves and/or
temperature measurements as diagnostics of photospheric inhomogeneities
and the H$\alpha$ and \ion{He}{1} D$_{3}$ lines as chromospheric diagnostics.

Only for $\kappa^1$ Cet, for which we obtained contemporaneous photometric 
and temperature rotational modulations, are we able to determine a unique 
solution for the spot parameters (area and temperature) following the 
method described in \cite{Frasca05}; for the other stars we give only 
a rough estimate of these spot parameters.  Moreover, we apply a spot/plage 
model to the photospheric and chromospheric rotational modulations in order 
to investigate the degree of spot/plage association in these mildly active 
stars, comparing the results with previous results obtained for very 
active binaries \citep{Fra94,Frasca05} and single stars \citep{Frasca00}.

\section{OBSERVATIONS}

\subsection{Target Stars}

We selected four early-G to early-K main sequence stars to be observed
spectroscopically.  We also have contemporaneous photometry for two of 
the stars ($\kappa^1$~Cet, $\chi^1$~Ori).  The other two stars (HD 166, 
$\epsilon$~Eri) lack contemporaneous photometry.  All four stars are
listed by HD number in Table \ref{tab:tbl-1}, along with each star's
name, $B-V$ color index, spectral type, $P_{\rm rot}$, the 
photometric comparison star (for the two stars observed photometrically), 
and the stellar templates we used for spectral subtraction.

A summary of spectroscopic and photometric observations is listed in Table \ref{tab:tbl-2}.

\subsection{Photometry}
The photometric observations of $\kappa^1$~Cet and $\chi^1$~Ori were
acquired with the T4 0.75~m Automatic Photoelectric Telescope (APT) at 
Fairborn Observatory in southern Arizona (USA).  The APT is equipped with 
an EMI 9124QB photomultiplier detector that measures stars sequentially
through Str\"{o}mgren $b$ and $y$ filters.  The observations are reduced
differentially and corrected for extinction with nightly extinction
coefficients and transformed to the Str\"{o}mgren system with yearly
mean transformation coefficients.  A complete discussion of the operation of
this telescope and the reduction of the resulting data can be found in 
\cite{Hen99}.  In this paper, we have analyzed data on $\kappa^1$~Cet 
and $\chi^1$~Ori acquired from November 2000 to January 2001, i.e., 
contemporaneous to the spectroscopic observations.

\subsection{Spectroscopy}
Spectroscopic observations have been performed in 2000 and 2001 at the 
{\it M. G. Fracastoro} station (Serra La Nave, Mt. Etna) of Catania 
Astrophysical Observatory with FRESCO (Fiber-optic Reosc Echelle 
Spectrograph of Catania Observatory).  The {\it \'echelle} spectrograph 
is connected to the 0.91~m telescope through a fiber link.  The spectral 
resolution was $R=\lambda/\Delta\lambda\,\simeq\,$14\,000, with a 2.6-pixel 
sampling.  The data reduction was performed with the {\sc echelle} task of 
the IRAF\footnote{IRAF is distributed by the National Optical Astronomy 
Observatory, which is operated by the Association of the Universities for 
Research in Astronomy, inc. (AURA) under cooperative agreement with the 
National Science Foundation.} package following the standard steps: 
background subtraction, division by a flat field spectrum from a 
halogen lamp, wavelength calibration using the emission lines of a 
Th-Ar lamp, and normalization to the continuum through a polynomial fit.  
Further details about the instrumentation and data reduction can be found 
in \cite{Cata02}.

Our spectra include the H$\alpha$-$\lambda$6563 and \ion{He}{1}-$\lambda$5876
lines and a number of photospheric lines used for the temperature 
determination described in Section \ref{sec:temp}. 

\section{TEMPERATURE AND H$\alpha$/\ion{He}{1} \\ANALYSIS}\label{sec:temp}

Temperature determinations of our target stars have been obtained by 
measuring the depth ratio of several line pairs following a method described 
by \cite{Cata02}.  The line-depth ratios (LDRs) allow us to 
resolve temperature variations as small as 10\,K \citep{Gray91,Gray01}, 
and the precision improves when one considers the average of several line 
pairs.  For example, \cite{Cata02} have demonstrated that LDRs 
can be used to detect the rotational modulation of the disk-averaged 
stellar temperature caused by the passage of cool spots across the 
photospheric disks of active RS~CVn stars.

The H$\alpha$ line has proven to be a very good diagnostic of stellar 
chromospheric activity and is easily accessible at optical wavelengths.  
Consequently, we have extracted the excess emission in the H$\alpha$ 
line that, in mildly active stars, partially fills the core of the 
H$\alpha$ absorption profile.  The emission contribution has been extracted
with the ``spectral synthesis" method.  High $S/N$ spectra of standard stars 
with negligible activity have been used as inactive templates for the 
spectral subtraction (see Table~\ref{tab:tbl-1}).  The convolution 
of the template spectra with a proper rotational profile to mimic the 
$v\sin i$ of each target was not necessary because the stars analyzed in 
this paper have rotational velocities lower than 7 km s$^{-1}$, which is about 
the FRESCO resolution.

As an additional diagnostic of the upper chromosphere, we have used the 
\ion{He}{1} $\lambda$5876 line which is seen as an absorption feature 
in the residual spectra.

\subsection{$\chi^1$~Ori = HD 39587} \label{sec:chi1_ori}

\objectname{$\chi^1$~Ori} ($V$=4\fm41) is a main sequence star that was 
first detected as an astrometric binary by \cite{Lippi78} and then discovered 
to be a long-period SB1 ($P_{\rm orb}$ = 5156$\fd$7) by \cite{Han02}.  
However, the presence of a low-mass companion in such a wide system should 
not affect the activity pattern of $\chi^1$~Ori.  From a long-term 
\ion{Ca}{2} H\&K chromospheric emission analysis, \cite{Baliu95} observed 
significant variability but with no clear period.  The star is a relatively 
rapid rotator since it is a young star belonging to the Ursa Major Cluster 
with an age of about 300 Myr.  \cite{Konig02} find $M_1=1.01$ $M_{\odot}$ and 
$M_2=0.15$ $M_{\odot}$ from an $H$-band image of the secondary component 
taken with the Keck adaptive optics system.  The \ion{He}{1} line was observed 
in $\chi^1$~Ori~A in absorption by \cite{Lambert83}.

\begin{figure}
\epsscale{1.1}
\plotone{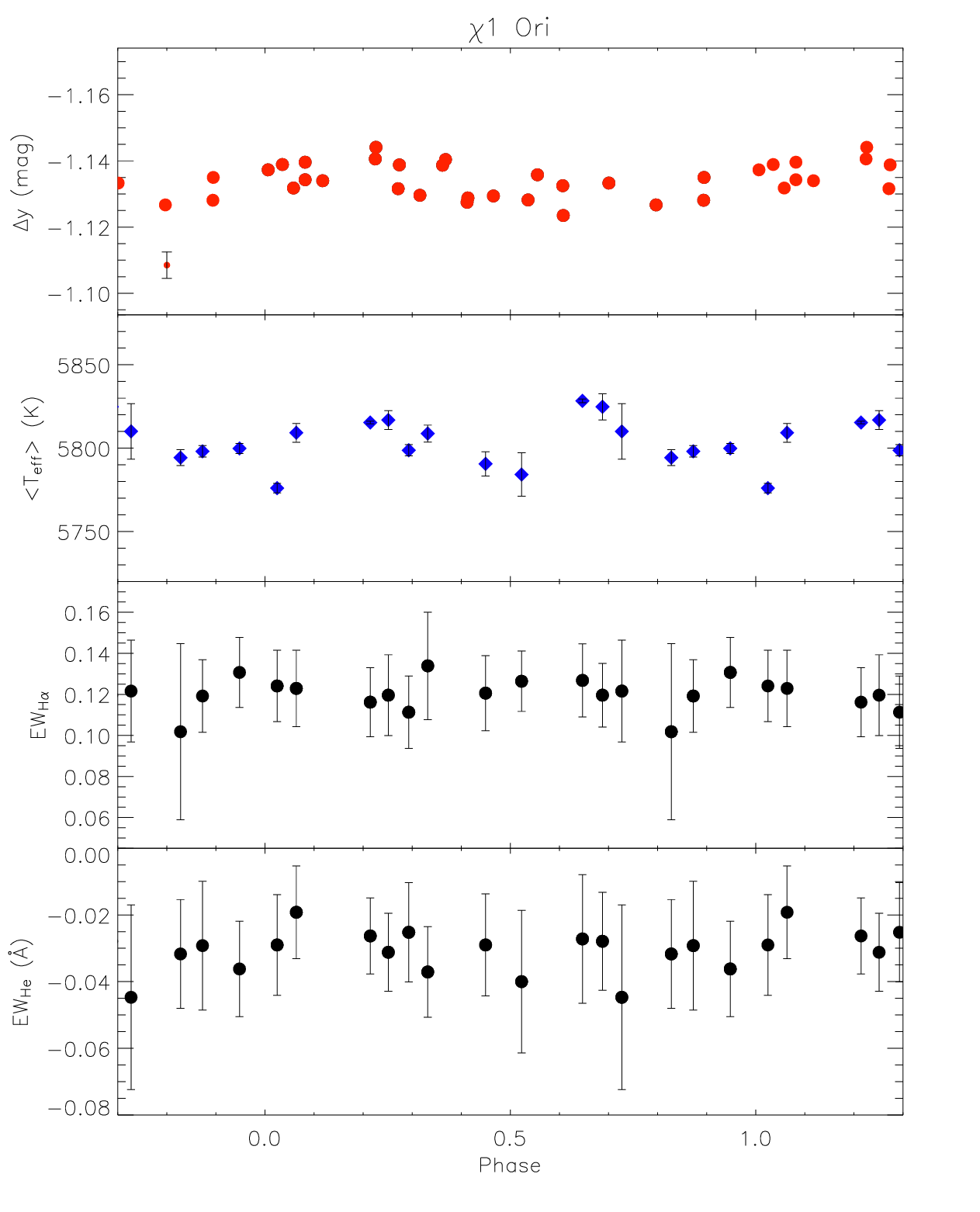}
\caption{{\it From top to bottom.} Differential Str\"{o}mgren $y$ photometry, 
$<T_{\rm eff}>$, $EW_{\rm H\alpha}$, and $EW_{\rm He}$ all ploted as a 
function of the rotational phase for $\chi^1$~Ori.}
\label{fig:f1}
\end{figure}

Contemporaneous temperature, light, and H$\alpha$ emission curves of this 
magnetically active star are plotted in Fig. \ref{fig:f1}.  The 
data have been folded in phase with the ephemeris $HJD_{\phi=0} = 
2\,451\,856.0 + 5\fd24\,\times\,E$, where the initial epoch is the date of
the first observation and the rotational period is from \cite{Messi01}.  
The $\Delta y$ photometry shows a low amplitude modulation of $\sim$0\fm02.  
However, neither the net H$\alpha$ equivalent width nor the derived 
temperature of $\chi^1$~Ori appear to exhibit rotational modulation
(Fig.~\ref{fig:f1}).  The maximum derived temperature is 
5828 K, which is close to the value of 5838 K in \cite{Gray94}, found 
by means of spectral line-depth ratios from high-resolution spectra.

\begin{figure*}
\epsscale{1.}  
\plotone{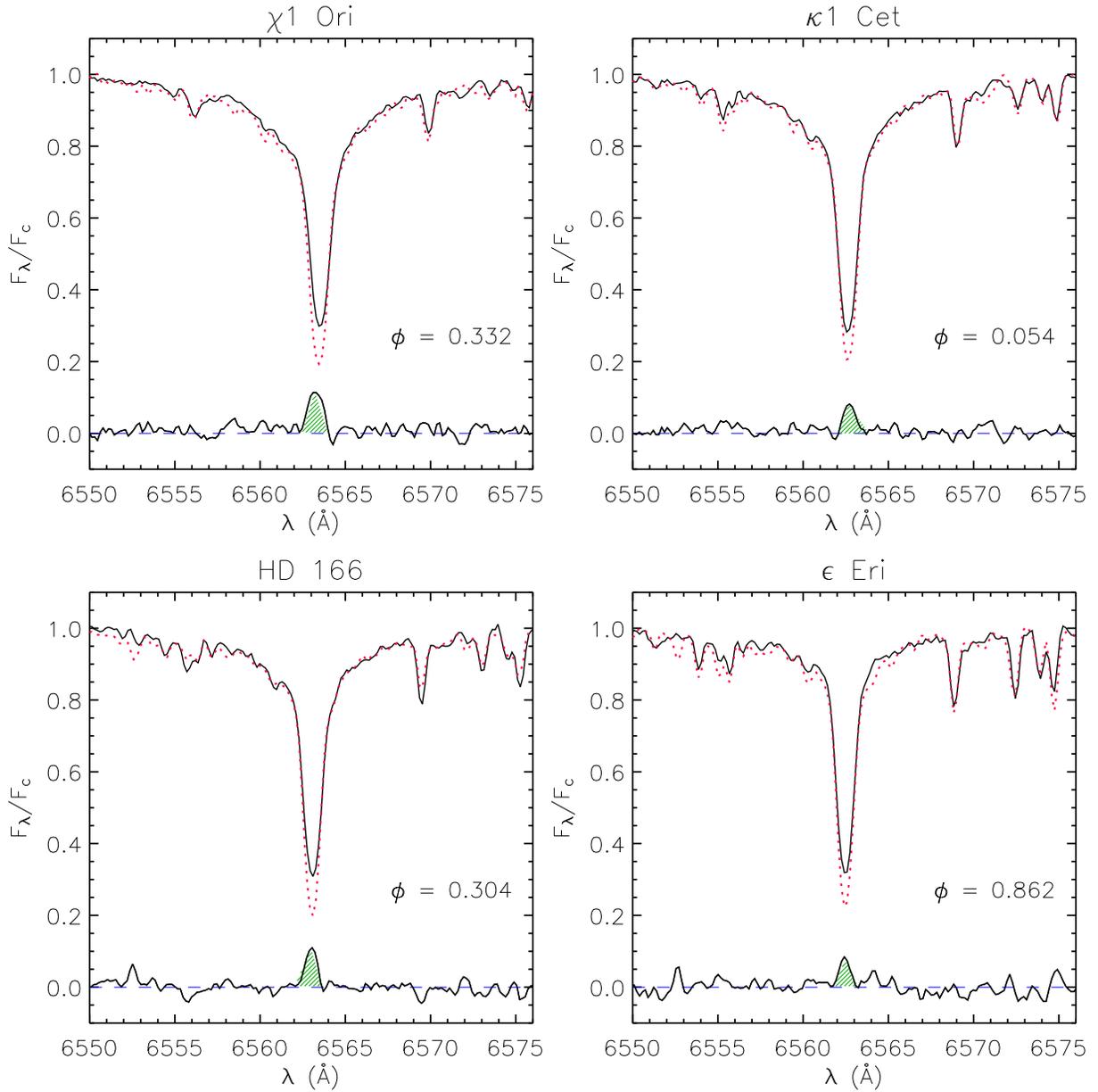}
\vspace{-.5cm}
\caption{{\it Top of each panel}: The observed, continuum-normalized spectra 
of the target stars (solid line) in the H$\alpha$ region together with the 
inactive stellar template (dotted line). {\it Bottom of each panel}: The 
difference spectra of the two upper spectra.}
\label{fig:f2}
\end{figure*}

Spectra of the H$\alpha$ line of all four stars in our sample are shown
in Fig.~\ref{fig:f2}.  The H$\alpha$ profile in 
$\chi^1$~Ori and the other three stars is always partially filled-in by
emission.  Since $\chi^1$ Ori is a rather active star, as denoted by its 
excess H$\alpha$ emission, it may be that the lack of rotational modulation
in temperature and H$\alpha$ emission is due to observations that were 
acquired at an epoch of relatively low activity or at a time when the active 
regions were evenly distributed in longitude.  This could also explain the 
very low amplitude of the light curve. 

The \ion{He}{1} line is also detected as an absorption feature in the 
spectra of $\chi^1$~Ori with values of equivalent width around 30 m\AA, but 
it does not show a modulation with the phase.  \cite{Lambert83} find a 
value of $EW_{\rm He}$=29 m\AA, and \cite{Danks85} obtain 
$EW_{\rm He}$= 25 m\AA, i.e. very close to the equivalent width measured 
in this work. 

\subsection{$\kappa^1$~Cet = HD 20630} \label{sec:k1_cet}

\objectname{$\kappa^1$~Cet} ($V$=4\fm83) is a nearby (9.16 pc) single G5 dwarf.  
Evidence of rotational modulation of \ion{Ca}{2} H\&K chromospheric emission 
has been found by \cite{Vau81}.  Changes in its photometric (rotational) 
period suggest a combination of differential rotation and concentration of 
starspots at different stellar latitudes from year to year 
(\citealt{Gaidos00}), consistent with a latitude drift of starspots during 
an activity cycle.  In fact, \cite{Messina02} find the existence of a 
solar-like starspot cycle of 5.9 years, which is similar to the chromospheric 
activity cycle of 5.6 years found by \cite{Baliu95}.  \cite{Gudel97} 
estimated an age of 750 Myr for $\kappa^1$~Cet from the relatively rapid 
rotation period of 9\fd2 seen in the spot modulation and 
suggested that the star is a likely member of the Hyades moving group.  
The \ion{He}{1} line was observed in absorption by \cite{Lambert83}, and 
its equivalent width in the $17.4-25.8$ m\AA~range appeared rotationally 
modulated.

\begin{figure}
\epsscale{1.1}
\plotone{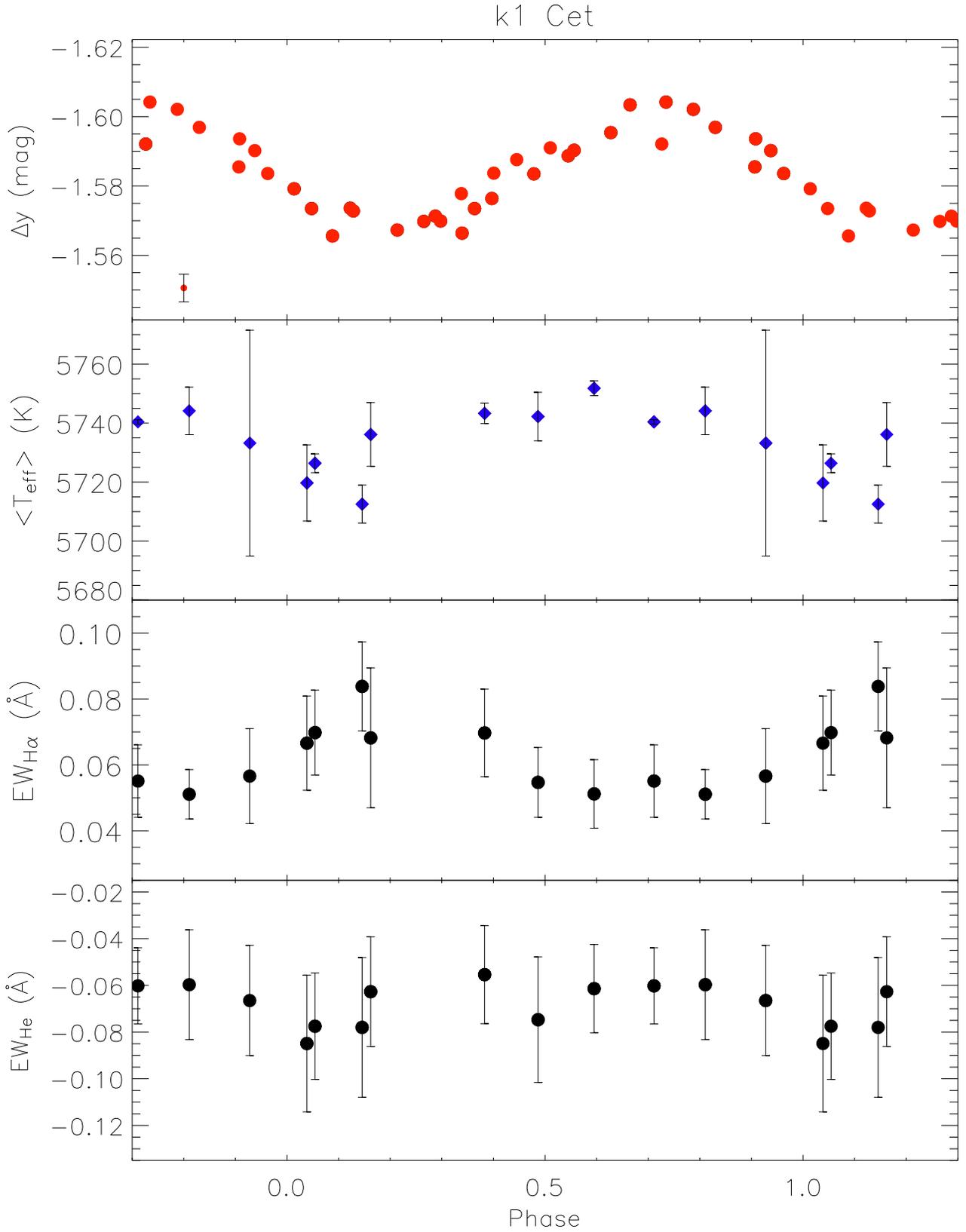}
\caption{{\it From top to bottom.} Differential Str\"{o}mgren $y$ photometry, 
$<T_{\rm eff}>$, $EW_{\rm H\alpha}$, and $EW_{\rm He}$ as a function of the 
rotational phase for $\kappa^1$~Cet.} 
\label{fig:f3}
\end{figure}

The temperature variation, derived from our spectra is shown in 
Fig.~\ref{fig:f3} together with the contemporaneous 
light curve.  The rotational phases have been computed from the ephemeris 
$HJD_{\phi=0} = 2\,451\,856.0+9\fd20\,\times\,E$, where the initial 
epoch is the date of the first observation and the mean rotational period 
is from \cite{Gaidos00}.  The two curves correlate fairly well, each with a
minimum around $\phi \simeq 0\fp15$ and a maximum near $\phi \simeq 0\fp65$.  
The amplitude of the temperature curve is about 40 K, with an average value 
of $\sim$5730~K, close to the values of 5718 K and 5747 K measured by \cite{Gray94} 
and \cite{Gaidos02} by means of spectroscopic analyses.  The light curve has 
an amplitude of about 0\fm04.

A spectrum of $\kappa^1$~Cet in the H$\alpha$ region is shown in 
Fig.~\ref{fig:f2}.  The core of the H$\alpha$ profile 
is always slightly filled-in by emission.  The net equivalent width, as measured 
in the residual spectra, and the average temperature values are plotted in 
Fig.~\ref{fig:f3}.  An anti-correlation between the light 
curve and $EW_{\rm H\alpha}$ modulation is apparent with the H$\alpha$ 
minimum at $\phi\simeq0\fp65$ and the H$\alpha$ maximum near $\phi=0\fp15$, 
i.e., at the same rotational phases as the maximum and the minimum of the 
light and temperature curves, respectively.  This implies a strong spatial 
correlation between the stellar spots and the chromospheric plages. 

The residual H$\alpha$ profile of $\kappa^1$~Cet is relatively narrow 
($FWHM=0.80-1.06$ \AA) and does not display the broad wings or asymmetric 
shapes observed in very active RS CVn stars.  This implyes that the 
chromospheric active regions in this solar-type star, which is more active 
than the Sun, nonetheless has a structure similar to the solar plages and 
lacks the strong mass motions and broadening effects observed in many 
of the more active RS~CVn stars (\citealt{Hatzes95,Biazzo06}). 

The \ion{He}{1} line is always observed as an absorption feature whose 
intensity varies slightly.  The large relative errors prevent us from
reliably establishing any correlation with the rotational period.  
The average value of our $EW_{\rm He}$ is about 70 m\AA, i.e., higher than 
previous results obtained by several authors \citep{Lambert83,Danks85,Saar97}.

\subsection{HD~166} \label{sec:hd166}

\objectname{HD~166} ($V$=6\fm13) is a nearby (13.7 pc) young solar-type 
star belonging to the Local Association, a young moving group with stars in 
an age range from about 50 to 150 Myr (\citealt{Montes01}).  HD~166 was 
first found to be a variable star by \cite{Rufe82}, who observed 
``microvariability'' in the star's $V$ magnitudes.  New photometric 
observations were presented by \cite{Gaidos00}; they found the star to vary 
with an amplitude up to 0\fm04 with a period of 6\fd23.  The \ion{He}{1} 
line was observed in absorption by \cite{Saar97}.

\begin{figure}
\epsscale{1.1}
\plotone{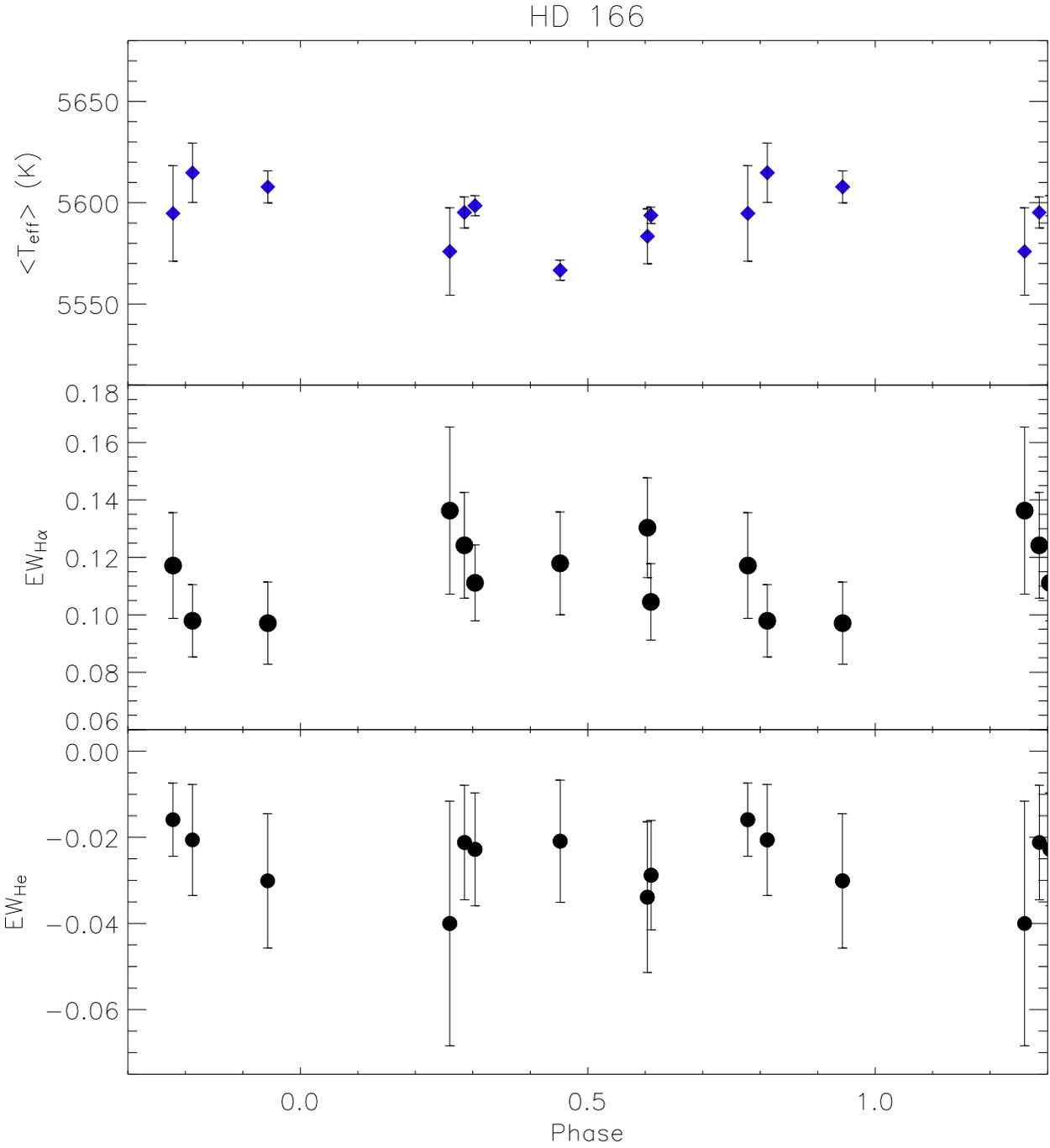}
\caption{{\it From top to bottom.} $<T_{\rm eff}>$, $EW_{\rm H\alpha}$ and 
$EW_{\rm He}$ as a function of the rotational phase for HD~166.
\label{fig:f4}}
\end{figure}

Fig.~\ref{fig:f4} plots the temperatures derived from the LDRs in 
FRESCO spectra as a function of the rotational phase, where the ephemeris 
used is from \cite{Gaidos00}: $HJD_{\phi=0} = 2\,449\,540.0 + 6\fd23\,
\times\,E$.  HD 166 shows a clear rotational modulation of the average 
temperature with an amplitude of 48 K and a maximum of 5615 K very close to 
the value of 5620 K found spectroscopically by \cite{Gaidos02}.  Unfortunately, 
for this star, no simultaneous light curve is available. 

The variation in the $EW_{\rm H\alpha}$ is also plotted as a function of the 
rotational phase in the same figure.  We used $\tau$~Cet, one of the stars 
with the lowest level of activity ever observed, as our template star.
Notwithstanding the scatter in the $EW_{\rm H\alpha}$ data, an anti-correlation 
between photospheric and chromospheric diagnostics is visible.  A spectrum 
of HD~166 in the H$\alpha$ region is presented in 
Fig.~\ref{fig:f2}, where the filling-in is evident.

The He equivalent width that we find has an average value of about 28 m\AA, 
that is close to the value of 20 m\AA~obtained by \cite{Saar97}.

\subsection{$\epsilon$ Eri = HD 22049} \label{sec:eps_eri}

\objectname{$\epsilon$~Eri} ($V$=3\fm73) is a bright, nearby (3.3 pc) 
single K2 main sequence star that shows variability attributed to magnetic 
activity.  From LDR analysis, \cite{Gray95} find a temperature excursion 
of about 15 K during the 1986--1992 time interval with rising temperatures 
associated with higher levels of magnetic activity in the cycle.  
Long-term photometry has been acquied by \cite{Frey91} and has led to the 
detection of a variable rotational period 10\fd0$<P_{\rm rot}<$12\fd3, 
indicative of latitude drift of starspots and differential rotation.  
\cite{Baliu95} measure the \ion{Ca}{2} H\&K chromospheric emission and 
report a significant variability with no clear period from the power spectrum 
analysis.  Moreover, the \ion{He}{1} line was observed in absorption by 
\cite{Lambert83} with no rotational modulation.

\begin{figure}
\epsscale{1.1}
\plotone{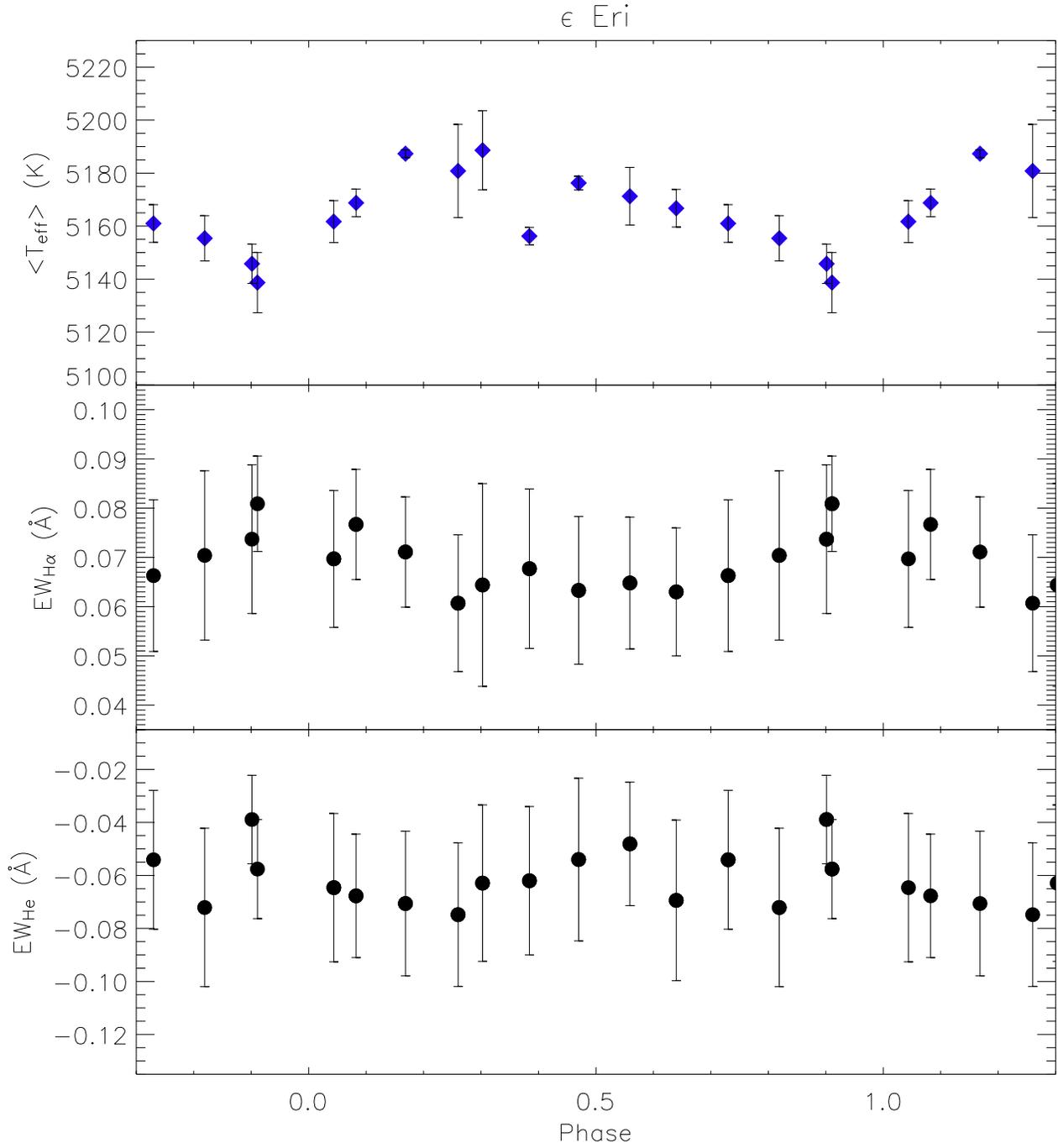}
\caption{{\it From top to bottom.} $<T_{\rm eff}>$, $EW_{\rm H\alpha}$ and 
$EW_{\rm He}$ as a function of the rotational phase for $\epsilon$~Eri.
\label{fig:f5}}
\end{figure}

Contemporaneous photometric data are not available for this star, thus only 
the analysis of the temperature variations has been done, as shown in 
Fig.~\ref{fig:f5}.  Phases have been computed from the ephemeris 
$HJD_{\phi=0} = 2\,451856.0+11\fd68\,\times\,E$, where the initial epoch 
is again the date of the first observartion and the rotation period is from 
the analysis of long-term chromospheric activity by \cite{Donahue96} at 
Mount Wilson.  A clear modulation of the disk-averaged temperature with 
rotational phase is apparent (Fig.~\ref{fig:f5}).  The average $T_{\rm eff}$ 
value (5164 K) is not far from the value of 5146 K found by \cite{Gray94} by 
means of spectral LDRs analysis.

A spectrum of $\epsilon$~Eri in the H$\alpha$ region is shown in 
Fig.~\ref{fig:f2}, while in the middle panel of Fig.~\ref{fig:f5} shows 
the results of the H$\alpha$ analysis.  A fairly well-defined anti-correlation 
is evident between the photospheric temperature curve and the net H$\alpha$ 
equivalent width curve.  The full amplitude of the temperature variation is 
50 K, i.e., about 1\%, while the net H$\alpha$ equivalent width excursion is 
about 33\% of its average value. 

The \ion{He}{1} line is also observed in absorption in the spectra of 
$\epsilon$~Eri with values of the disk-averaged equivalent width 
of about 55 m\AA, but it does not appear to be rotationally modulated.  
\cite{Lambert83}, \cite{Wolff84} and \cite{Danks85} find for 
this parameter values in the range 14$-$18 m\AA.

\section{SPOT/PLAGE MODELING}

\subsection{$\kappa^1$~Cet}

Following the arguments treated in \cite{Frasca05}, from the unspotted 
magnitude, $V_{\rm max} = 4\fm80$, and color index $(B-V)=0\fm68$ 
(\citealt{Messina02}), the stellar radius derived for $\kappa^1$~Cet is 
$R_{\rm 1} = 1.00$ $R_{\odot}$.  The temperature of the ``quiet'' photosphere 
is $T_{\rm ph}$ = 5752 K, while the derived inclination is 
$i = 55\degr_{-14}^{+28}$.

\begin{figure}
\epsscale{1.1}
\plotone{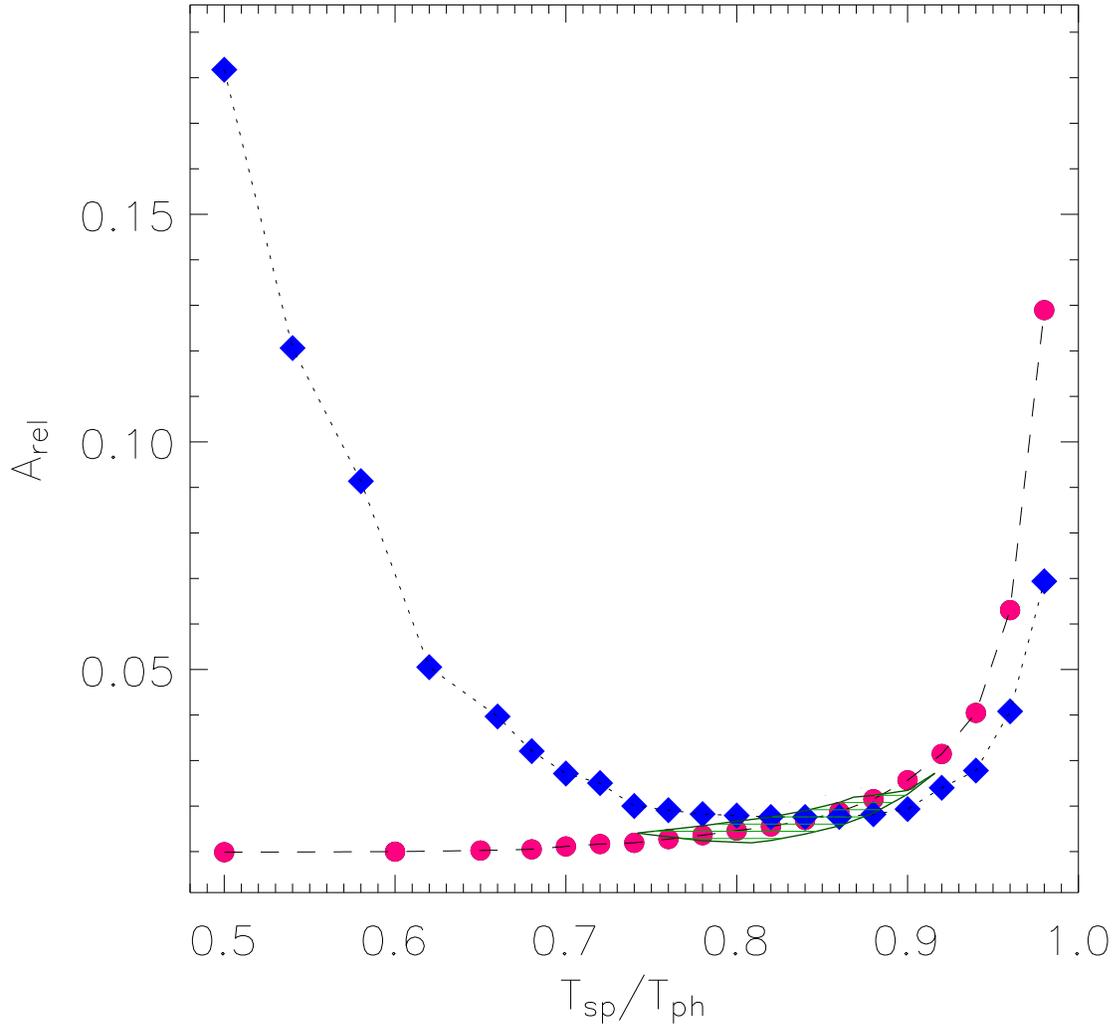}
\vspace{-1.cm}
\caption{Grids of solutions for $\kappa^1$~Cet. The filled circles represent 
the solutions for light curve, while the filled diamonds represent the solutions 
for temperature curve. The hatched area is the locus of the allowed solutions 
accounting for data errors.}
\label{fig:f6}
\end{figure}

As a consequence, applying the spot modeling developed by \cite{Frasca05}, 
which assumes two dark circular spots on a spherical limb-darkened star, 
we find two grids of solutions for the $\Delta y$ and $<T_{\rm eff}>$ curves, 
whose unique intersection provides the best values of the spot temperature 
$T_{\rm sp}$ and the projected area of the spots relative to the 
stellar surface $A_{\rm rel}$ (Fig.~\ref{fig:f6}).  We find a 
relative spot temperature $T_{\rm sp}/T_{\rm ph}$ = 0.847 and a 
relative spot area $A_{\rm rel}$ = 0.018, computing the flux ratio 
between spot area and quiet photosphere by means of synthetic ATLAS9 
(\citealt{Kurucz93}) low-resolution spectra. 

For the H$\alpha$ curve, a simple ``plage'' model with two bright plages has 
been applied (\citealt{Frasca00}).  We have fixed the emission flux ratio 
between plages and quiet chromosphere $F_{\rm plage}/F_{\rm chom}$=3, 
that is near to the typical value of the brightest solar plages.

In Table~\ref{tab:tbl-3} the spot/plage configuration derived 
from the model is reported, where $\mu_{\rm y}$ and $\mu_{6200}$ are the 
linear limb-darkening coefficients for the $y$ band and for the continuum 
at 6250 \AA.  $EW_{\rm chrom}$ is the value of the H$\alpha$ equivalent width 
at the maximum of the rotational modulation.

\begin{figure}
\epsscale{1.2}
\plotone{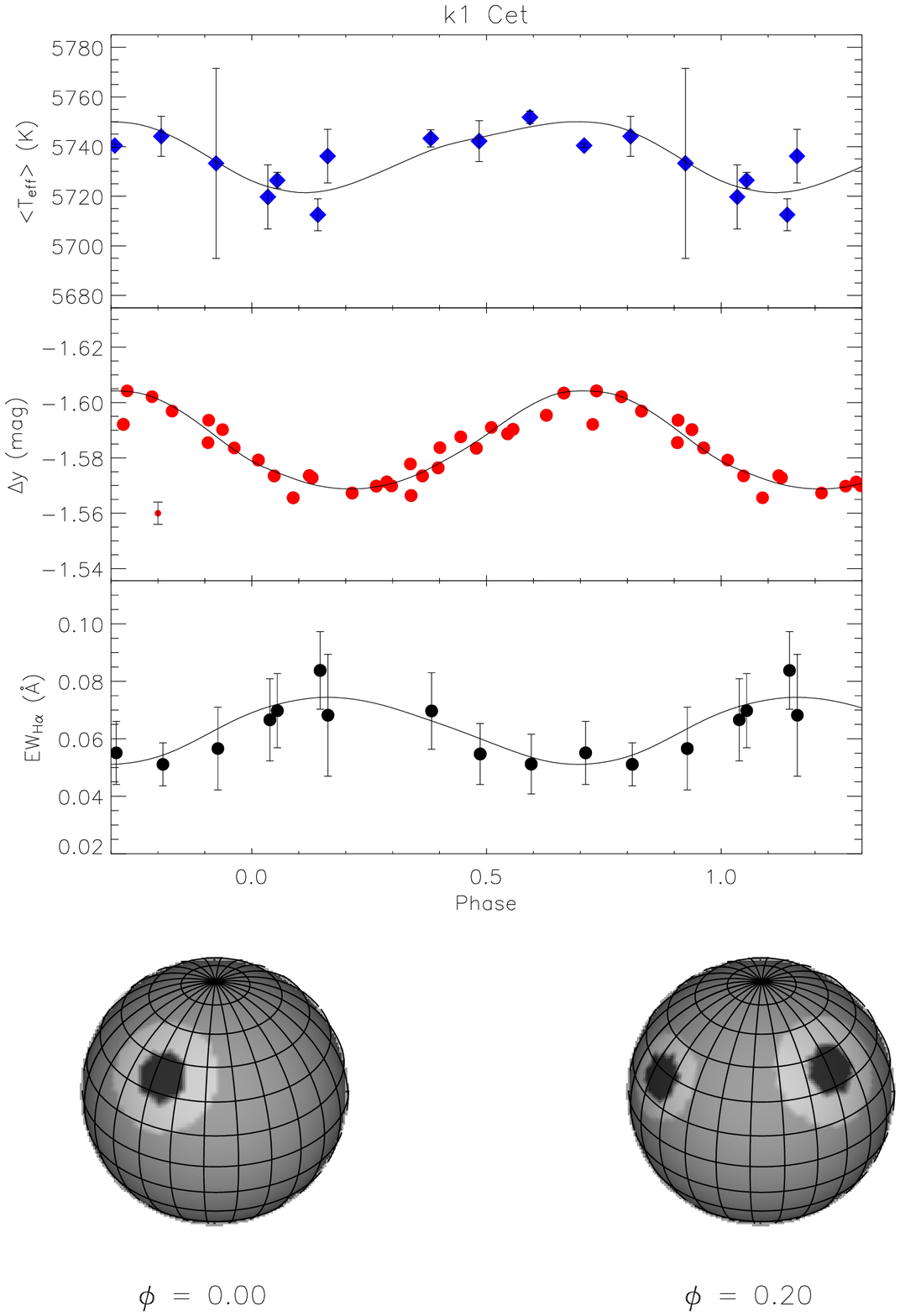}
\caption{Observed (filled circles) and synthetic (solid lines) temperature, 
light and $EW_{\rm H\alpha}$ curves of $\kappa^1$~Cet. The solutions are those 
obtained with the Kurucz model (\citealt{Kurucz93}).}
\label{fig:f7}
\end{figure}

The photospheric and chromospheric active regions have no appreciable 
longitude shifts (Table~\ref{tab:tbl-3}, Fig.~\ref{fig:f7}).  This result is similar 
to that obtained by \cite{Frasca00} for the young solar type star HD 206860.

\subsection{HD~166 and $\epsilon$~Eri}

For HD 166 and $\epsilon$~Eri we have no photometric data contemporaneous 
to the spectroscopic data.  As a consequence, we have applied the spot 
modelling only to the temperature curve, obtaining the minima of the 
temperature grids for $A_{\rm rel}=0.021$ and 0.026, which correspond to 
$T_{\rm sp}/T_{\rm ph}=0.84$ and 0.86, respectively, for HD~166 
and $\epsilon$~Eri.  In Table \ref{tab:tbl-3} we list the values 
of the approximate spot solutions.  These values are listed without error 
bars because it is impossible with only the temperature curve to define 
the locus of the allowed solutions.  Because in the cases of $\kappa^1$ Cet 
here presented and of the RS CVn binaries analyzed by \cite{Frasca05}, 
the unique spot solution obtained from the intersection of the two light 
and temperature grids is near the minimum of the grid of solutions for 
the temperature curve, we assume that, for HD~166 and $\epsilon$~Eri, the 
unique solution is near the temperature grid minima of 
$T_{\rm sp}/T_{\rm ph}=0.84$ and 0.86, respectively. 

Finally, the simple plage model described by \cite{Frasca00} has been 
applied to these two stars, fixing $F_{\rm plage}/F_{\rm chrom}=3$.  
The derived plage parameters are recorded in Table \ref{tab:tbl-3}.

\begin{figure*}
\epsscale{1.5}  
\plottwo{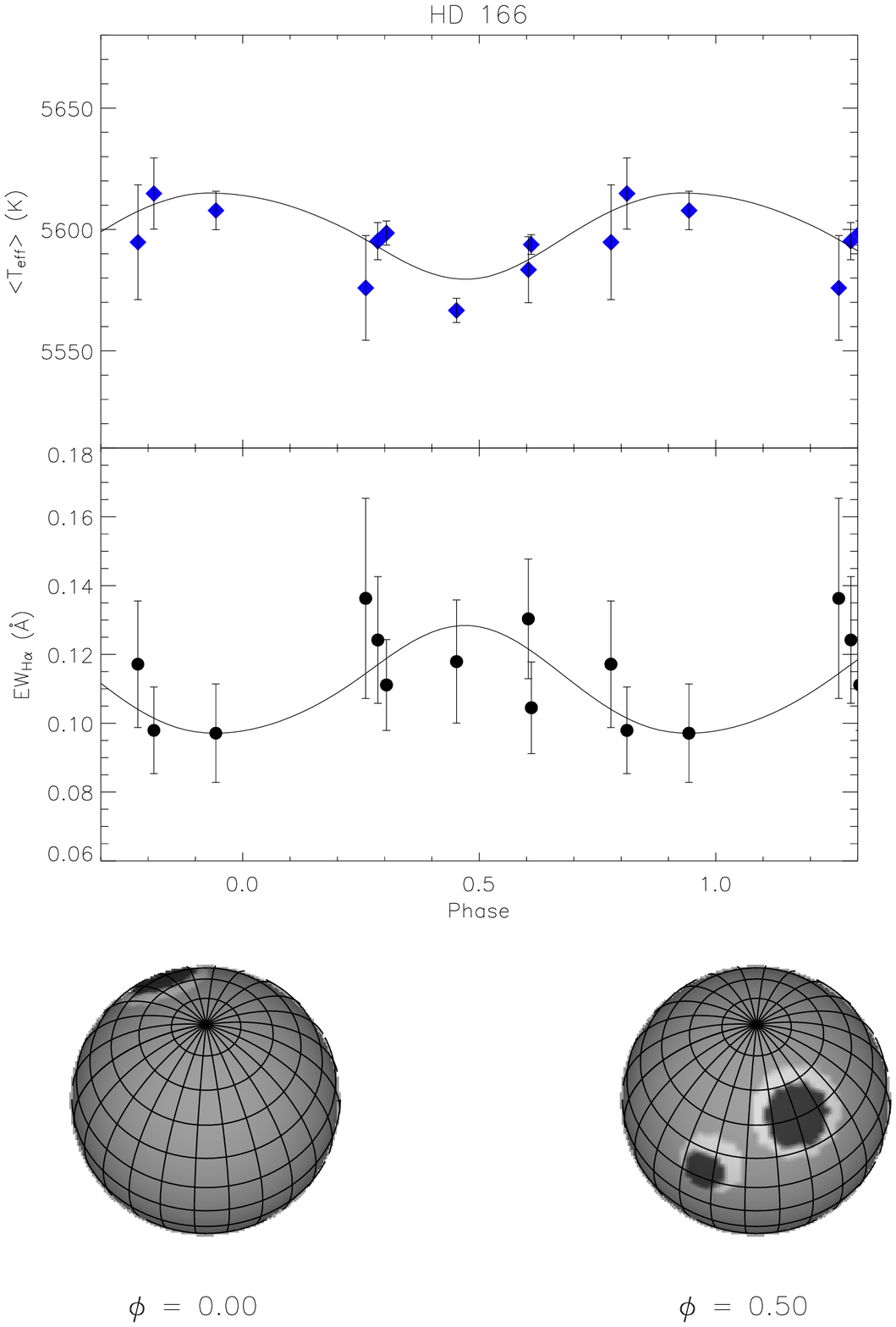}{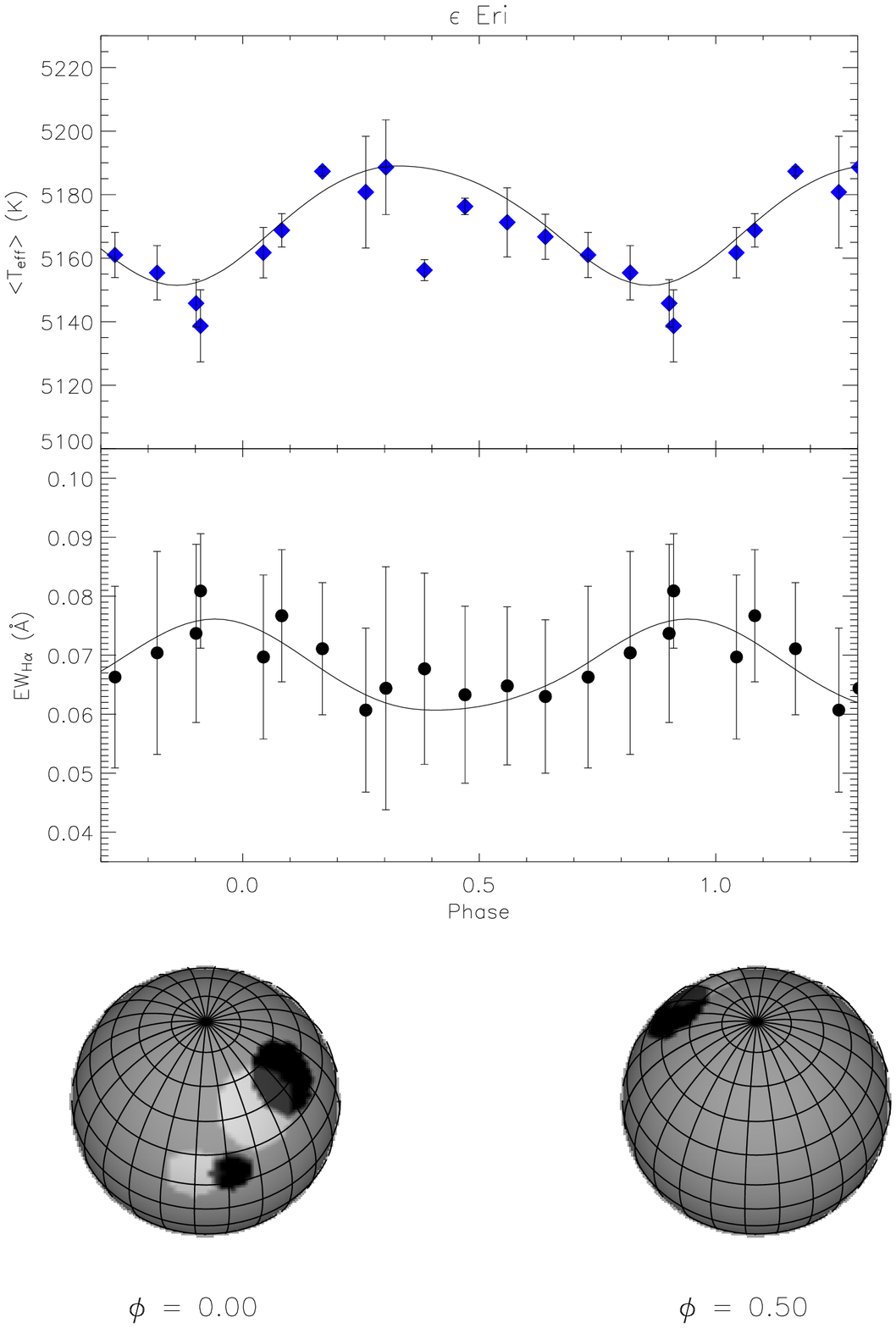}
\caption{Observed (filled circles) and synthetic (solid lines) temperature, 
$EW_{\rm H\alpha}$, and $EW_{\rm He}$ curves of HD~166 and $\epsilon$~Eri.}
\label{fig:f8}
\end{figure*}

The spot/plage configuration of HD~166 and $\epsilon$~Eri is displayed in 
Fig.~\ref{fig:f8}.

\section{Conclusion}

We have analyzed the photospheric and chromospheric activity in four young,
magnetically active solar-type stars, namely $\chi^1$ Ori, $\kappa^1$ Cet, 
HD~166 and $\epsilon$ Eri.  The photospheric surface features have been 
recovered by means of the rotational modulation of luminosity and temperature 
as derived from the LDR method, while the chromospheric inhomogeneities have 
been studied from their excess H$\alpha$ emission.  The H$\alpha$ profiles 
can be reasonably well reproduced by means of only one Gaussian component, 
indicating the presence of plages spatially associated with the photospheric 
spots.  The temperature and light curves are always anti-correlated with 
the H$\alpha$ emission modulation, confirming a close spatial association 
between spots and plages.  The only exception is $\chi$1 Ori, for which we 
have not obtained clear rotational modulation of temperature and H$\alpha$ 
curves.  The He line is always present in our spectra, but the $EW_{\rm He}$ 
measurements show too much scatter to reveal any modulation with the 
rotational phase.  Thus, in general, the active regions in mildly active 
stars seem to have structures similar to solar active regions. 

The spectroscopic measurements of HD~166 and $\epsilon$ Eri span about 
five stellar rotations, but the spot/plage configuration seems to be largely 
unchanged, as observed in some other young solar analogues where the light 
curve remains stable for several rotations (\citealt{Messina02}).

Moreover, from a simple spot/plage model analysis, we have derived the spot 
and plage parameters (temperature and area).  In the case of $\kappa^1$ Cet, 
for which we had both simultaneous photometric and spectroscopic data, spot 
temperature and area have been uniquely determined.  The grid of the temperature 
solutions is flat caused by the small-amplitude of the temperature curve 
of this MS star, and this leads to great errors in the unique spot solution.  
For HD~166 and $\epsilon$ Eri, we have also presented a rough estimate of these two 
parameters.  

Finally, the temperature difference $\Delta T$ between the quiet photosphere and spots, 
a key parameter tied to the blocking effect on convection produced by the 
intensification of the magnetic field, seems to be higher compared to that 
derived by \cite{Frasca05} for the stars with lower gravity, such as IM Peg 
($\Delta T =$ 448 K) and HK Lac ($\Delta T =$ 767 K).  On the other hand, the 
spot filling factors seem to be smaller than those obtained for these active 
stars.  Moreover, the chromospheric plages are larger than the associated spots, 
as observed in the Sun.  As a consequence, the increasing dissipation of magnetic 
energy with height above the photosphere, a characteristic of the Sun, seems to 
occur for these solar-type stars as well. 

In future studies, we want first to extend the number of main sequence targets 
and then investigate the inhomogeneities at photospheric and chromospheric 
levels in pre-main sequence, rapidly rotating stars to separate gravitational 
effects from activity level effects.

\acknowledgments
This work has been supported by the Italian {\em Ministero dell'Istruzione, 
Universit\`a e  Ricerca} (MIUR) and by the {\em Regione Sicilia}, which are 
gratefully acknowledged.  GWH acknowledges support from NASA grant NCC5-511
and NSF grant HRD-9550561.  This research has also made use of SIMBAD and 
VIZIER databases, operated at CDS, Strasbourg, France.

\clearpage

\begin{table*}
\begin{center}
\caption{STELLAR SAMPLE\label{tab:tbl-1}}
\begin{tabular}{rcccrrr}
\tableline
\tableline
HD Number & Name & $B-V$ & Sp. Type & $P_{\rm rot}^\dagger$ & Comp. Stars & Spectr. Templates\\
 &  & (mag) &  & (d) &  & \\
\tableline
166    & ...            & 0.750 & K0V & 6.23 &  --	    &$\tau$~Cet\\
20630  & $\kappa^1$~Cet  & 0.680 & G5V & 9.20 &  HD 21585 &51~Peg\\
22049  & $\epsilon$~Eri & 0.880 & K2V & 11.68 &  --	    &54 Psc\\
39587  & $\chi^1$~Ori    & 0.590 & G0V & 5.24 & HD 37147 &10 Tau\\
\tableline
\end{tabular}
\end{center}
$^\dagger$ For references see the text.
\end{table*}

\clearpage

\begin{table*}
\begin{center}
\caption{SUMMARY OF OBSERVATIONS\label{tab:tbl-2}}
\begin{tabular}{rccccrr}
\tableline
\tableline
HD Number & Spectr. data range & N$_{\rm obs}^{\rm spectr}$& Photom. data range& N$_{\rm obs}^{\rm photom}$\\
 & (JD -- 2\,400\,000) &  & (JD -- 2\,400\,000) &  \\
\tableline
166    & 51834.4--51865.4 & 9  & -- & -- \\
20630  & 51856.5--51866.5 & 10 & 51810.0--51975.6 & 44 \\
22049  & 51856.5--51917.4 & 13 & -- & -- \\
39587  & 51913.4--51867.6 & 14 & 51857.9--51913.7 & 24 \\
\tableline
\end{tabular}
\end{center}
\end{table*}

\clearpage

\begin{table*}  
\caption{SPOT/PLAGE CONFIGURATION FOR $\kappa^1$~Cet, HD 166, and $\epsilon$ Eri.}
\label{tab:tbl-3}
\begin{center}
\begin{tabular}{crccccc}
\hline
\hline
Radius  &  Lon.$^{a}$  &   Lat.  &  $T_{\rm sp}/T_{\rm ph}$ &  $T_{\rm ph}$ &  $T_{\rm sp}$ &  $A_{\rm rel}^{b}$ \\ 
\hline
~\\
\multicolumn{7}{c}{$\kappa$1 Cet}\\
\multicolumn{7}{c}{($\mu_{\rm y}$ = 0.669, $\mu_{6200}$ = 0.56, $EW_{\rm chrom}$=0.051 \AA)}\\
\hline
\multicolumn{7}{c}{S{\sc pots}}\\
\hline
12$\fdg$2   &  30$\degr$  & 38$\degr$  & 0.848$^{+0.068}_{-0.106}$ & 5752 K & 4878$^{+391}_{-610}$ K & 0.018$^{+0.009}_{-0.006}$ \\
 9$\fdg$0   & 130$\degr$  & 28$\degr$  &  & & &	    \\
\hline
\multicolumn{7}{c}{P{\sc lages}}\\
\hline
24$\fdg$8   &  211$\degr$  &142$\degr$    &   &  & & 0.074 \\
19$\fdg$3   &  310$\degr$  &152$\degr$    &   &  & &  \\
\hline
~\\
\multicolumn{7}{c}{HD~166}\\
\multicolumn{7}{c}{($\mu_{6200}$ = 0.56, $EW_{\rm chrom}$=0.097 \AA)}\\
\hline
\multicolumn{7}{c}{S{\sc pots}}\\
\hline
14$\fdg$1   & 153$\degr$  & 46$\degr$  & 5615 K & 0.840      & 4717 K  & 0.021 \\
 9$\fdg$0   & 204$\degr$  & 20$\degr$  & &     &   &  \\
\hline
\multicolumn{7}{c}{P{\sc lages}}\\
\hline
20$\fdg$5   & 152$\degr$  & 47$\degr$	 & &   &  & 0.047 \\
14$\fdg$2   & 201$\degr$  & 24$\degr$	 & &   &  &  \\
\hline
~\\
\multicolumn{7}{c}{$\epsilon$ Eri}\\
\multicolumn{7}{c}{($\mu_{6200}$ = 0.59, $EW_{\rm chrom}$=0.061 \AA)}\\
\hline
\multicolumn{7}{c}{S{\sc pots}}\\
\hline
16$\fdg$3   & 298$\degr$  & 48$\degr$  & 5189 K & 0.860      & 4463 K  & 0.026 \\
 9$\fdg$1   & 348$\degr$  & 21$\degr$  & &     &   &  \\
\hline
\multicolumn{7}{c}{P{\sc lages}}\\
\hline
17$\fdg$8   & 326$\degr$  & 46$\degr$	&  &   &  & 0.034 \\
11$\fdg$6   &	5$\degr$  & 21$\degr$	&  &   &  &  \\
\hline
\end{tabular}
\end{center}
~\\
{\small $^{a}$ Longitude increases with phase, and 0$\degr$ longitude
corresponds to phase 0$\fp$0.\\
$^{b}$ Total fractional area of the two spots. }
\end{table*}

\end{document}